\documentstyle[pra,manuscript,aps]{revtex}
\draft
\tighten
\begin{document} 
\title{Modified Double Exchange Model with Novel Spin
and Orbital Coupling: Phase Diagram of The Manganites} 
\author{L. Sheng and C.S. Ting} 
\address{Physics Department and Texas Center for Superconductivity, 
University of Houston, Houston, TX 77204} 
\maketitle 
\begin{abstract} 
From a general model of the Mn oxides
$R_{1-x}A_x$MnO$_{3}$, we derive 
an effective Hamiltonian in the low-energy subspace
using the projection operator method, in which
a novel coupling between the spin and orbital
degrees of freedom is included. A phase diagram for 
temperature $T$ versus doping concentration $x$
is computed by means of
Monte Carlo simulation. Our result is consistent
with experimental observations in the Mn oxides
with relatively wide conduction band, such as
Pr$_{1-x}$Sr$_x$MnO$_3$ and La$_{1-x}$Sr$_{x}$MnO$_3$. 
According to the obtained orbital ordering, we also predict
that the motion of charge carriers transforms from
three-dimensional to two-dimensional as $x$ is increased
beyond a critical value. 
\end{abstract} 
\date{}
\mbox{}\\ 
 
The peculiar magnetic and transport properties including 
the colossal magnetoresistance (CMR) of the manganites
$R_{1-x}A_x$MnO$_3$ (where $R=$La, Nd or Pr  
and $A=$Ca, Sr, Ba or Pb) have become the main focus of 
many recent research activities. In undoped $R$MnO$_3$,
each Mn atom has four outshell $3d$ electrons, three localized
$t_{2g}$ electrons forming an $S=3/2$ core spin, and
one itinerant electron filling into two degenerate $e_g$
orbitals, namely, $|+\rangle=d_{3z^2-r^2}$
and $|-\rangle=d_{x^2-y^2}$. $R$MnO$_3$ is usually an 
A-type antiferromagnetic (A-AF) insulator~\cite{b1,b2},
where the local spins are aligned ferromagnetically
in layers parallel to the $ab$ plane and antiferromagnetically
along the $c$ axis. Hole doping by 
partial substitution of $R^{3+}$ atoms by $A^{2+}$ atoms
soon destroys the A-AF order, and 
strong ferromagnetism occurs for $0.1\stackrel{<}{\sim}
x\stackrel{<}{\sim}0.4$~\cite{b3,b4,b5,bb5}.

Near $x=0.5$, the manganites exhibit
different magnetic structures, as controlled by
the conduction band width $W$. 
some manganites with relatively large $W$,
such as Pr$_{0.5}$Sr$_{0.5}$MnO$_3$~\cite{b5,b6,b7}, 
are in an A-AF state below  
a critical temperature $T_N$ and changes to a
ferromagnetic (FM) state above $T_N$ through a discontinuous
phase transition. Some manganites with 
small $W$, e.g., Pr$_{0.5}$Ca$_{0.5}$MnO$_3$~\cite{b8}
and La$_{0.5}$Ca$_{0.5}$MnO$_3$~\cite{bb8},
exhibit a more complicated CE-type antiferromagnetic 
(CE-AF) order with charge ordering at low temperatures, 
where the spins arrange ferromagnetically
in zig-zag chains and antiferromagnetically between
neighboring chains.
In some other manganites with intermediate $W$, such
as Nd$_{1-x}$Sr$_{x}$MnO$_3$~\cite{b7,b9}, CE-AF
and A-AF orders may coexist in the systems,
and their volume fractions change 
rapidly with small change in $x$. 

For further doping, the A-AF state appears to 
be the fundamental ground state spin configuration 
in many manganites.
It has been observed in La$_{1-x}$Sr$_{x}$MnO$_3$
with $0.52<x<0.58$~\cite{b10}, Nd$_{1-x}$Sr$_{x}$MnO$_{3}$
with $0.53<x<0.62$~\cite{b10} and Pr$_{1-x}$Sr$_{x}$MnO$_{3}$
with $0.48<x<0.6$~\cite{b5}. 
Near the ending doping concentration $x=1.0$,
the manganites develop an isotropic G-type
antiferromagnetic (G-AF) ground  
state~\cite{b1}. In the region
between $x\simeq 0.6$ and the ending concentration, most
manganites are non-ferromagnetic insulators~\cite{b4,b10}, 
but their microscopic spin structures have so far been
seldom studied in experiments.

The conventional double
exchange (DE) model~\cite{b11,b12,b13} 
for the Mn oxides usually predicts strong
ferromagnetism symmetrically about $x=0.5$,
which does not account for the  antiferromagnetic
phase at $x\stackrel{>}{\sim}0.5$
observed experimentally. While
Jahn-Teller (J-T) effect~\cite{b14,b15},
orbital degrees of freedom~\cite{b16}, and Coulomb
interactions~\cite{b5,b16} have been proposed 
as necessary extensions of the DE model,  
theoretical phase diagram in the $T\sim x$ plane
commensurating with experimental observations
has yet to be obtained.

In this work, starting from a general model
for the manganites we derive a modified
DE Hamiltonian, which by comparison with
the conventional DE model contains an orbital
dependence in the hopping integral and 
a novel effective coupling between the spin
and orbital degrees of freedom originating from the 
strong correlations. The magnetic and orbital phase diagram of 
this Hamiltonian is calculated using Monte Carlo simulation
and simulated annealing technique~\cite{b18}. Our results obtained are in 
good agreement with the experimental measurements
in the manganites with relatively wide conduction band 
in the entire $T\sim x$ plane. 

With the two-fold degeneracy of the $e_g$ orbitals
included, the total Hamiltonian for the Mn oxides 
can be written as
\begin{eqnarray}
H&=&-\sum\limits_{ij\alpha\beta}
t_{ij}^{\alpha\beta}a^\dagger_{i\alpha s}
a_{j\beta s}-K\sum\limits_{i\alpha ss'}
\mbox{\boldmath{$\sigma$}}_{ss'}\cdot{\bf s}_i a_{i\alpha s}^\dagger
a_{i\alpha s'}+U\sum\limits_{i\alpha}n_{i\alpha\uparrow}
n_{i\alpha\downarrow}\nonumber\\
&+&U'\sum\limits_{iss'}n_{i+s}n_{i-s'}
+\frac{J_{\mbox{s}}}{2}\sum\limits_{ij}
{\bf s}_i\cdot{\bf s}_j
-g\sum\limits_{i\alpha\beta s}
\mbox{\boldmath{$\tau$}}_{\alpha\beta}\cdot{\bf Q}_i
a_{i\alpha s}^\dagger a_{i\beta s}
+\frac{k}{2}\sum\limits_iQ_i^2\ .
\end{eqnarray}
Here, $a_{i\alpha s}$ annihilates an electron at site
$i$ in $e_g$ orbital $\alpha$$(=+$, or $-$) with spin
$s$$(=\uparrow$, or $\downarrow$).  
$K$ is the Hund's rule coupling between $e_g$ 
electrons and localized spins, where $\mbox{\boldmath{$\sigma$}}$
are the Pauli matrices in the $e_g$  
spin space and ${\bf s}_i={\bf S}_i/S$
a unit vector in the direction of the classical 
localized spin ${\bf S}_i$.  $U$ and $U'$ represent
the intraorbital and interorbital Coulomb repulsions,
respectively. $J_{\mbox{s}}$ stands for the superexchange
coupling of the localized $t_{2g}$ spins. $g$ is the J-T
coupling between the $e_g$ electrons and local 
J-T lattice distortions ${\bf Q}_i$,
where $\mbox{\boldmath{$\tau$}}$
are the Pauli matrices in the orbital space and ${\bf Q}_i=
(Q_{ix},Q_{iz})$ describes the two $e_g$ modes of the J-T
distortions. The orbital degrees of
freedom are also called the isospin of the $e_g$ electrons,
for they are similar to the spin degrees of freedom. 
The orbital-dependent
hopping integrals~\cite{b16} $t_{ij}^{\alpha\beta}$
are elements of the $2\times 2$ matrix $
\hat{t}_{ij}=t(1+\mbox{\boldmath{$\tau$}}
\cdot{\bf n}_{ij})$. 
Here, ${\bf n}_{ij}={\bf n}_\delta$
with $\delta=x,y,$ and $z$ for hopping along the $x$, $y$ and
$z$ direction, where ${\bf n}_x=(-\sqrt{3}/2,0,-1/2)$, ${\bf n}_y
=(\sqrt{3}/2,0,-1/2)$ and ${\bf n}_z=(0,0,1)$ are three
unit vectors distributed
symmetrically in the $x-z$ plane. 
Introducing the three unit vectors is a key
step in our theory, which will allow us to express the 
effective Hamiltonian in a compact
and transparent form. 
 
In the case of large $U$, $U'$, $K$ and 
J-T coupling $g$,
we can derive an effective Hamiltonian in the low-energy
subspace by following a similar procedure to the derivation of 
the $t-J$ model from the Hubbard model~\cite{b17}. 
The total Fock space is divided into two subspaces:
the low-energy subspace where 
each Mn site is occupied by at most one $e_g$ electron
and the spin and isospin of the $e_g$ electron
are parallel to local ${\bf s}_i$ and ${\bf Q}_i$,
respectively, and the reminder subspace, namely, the 
high-energy subspace. 
Using the projection operator method and treating the kinetic
energy term as a perturbation~\cite{b17,bb17}, we obtain
an effective Hamiltonian in the low-energy subspace
\begin{eqnarray}
H_{eff}&=&-\sum\limits_{ij}\widetilde{t}_{ij}d_{i}^\dagger
d_{j}+\frac{J_{\mbox{s}}}{2}\sum\limits_{ij}(1+\gamma n_in_j)
{\bf s}_i\cdot{\bf s}_j\nonumber\\
&-&\frac{J'}{2}\sum\limits_{ij}n_in_j
(1+\lambda{\bf s}_i\cdot{\bf s}_j)
\bigl[1-(\mbox{\boldmath{$\tau$}}_i\cdot{\bf n}_{ij})
(\mbox{\boldmath{$\tau$}}_j\cdot{\bf n}_{ij})\bigr]\ ,
\end{eqnarray}
where $n_i=d_i^\dagger d_i$,
and $d_i$ annihilates an $e_g$ electron 
with spin and isospin parallel to ${\bf s}_i$
and ${\bf Q}_i$. 
Here, $\gamma=2I_3/J_{\mbox{s}}$, $J'=I_1+I_2-I_3$,
and $\lambda=(I_1-I_2+I_3)/J'$, where
$I_1=t^2/(U'+2n_eE_J)$, $I_2=t^2/(U'+2n_eE_J+2K)$ 
and $I_3=t^2/(U+2K)$ with $E_J=g^2/k$ as the 
$J-T$ energy and $n_e=1-x$ as the average
electron density. The isospin and lattice distortion
are forced mutually parallel by the strong J-T coupling.
$\mbox{\boldmath{$\tau$}}_i$
is a unit vector describing the orientation  
of the isospin or which orbital state
on Mn site $i$ is occupied by an $e_g$ electron. 
If $\mbox{\boldmath{$\tau$}}_i$ makes an angle $\phi_i$
with the $z$ axis, then the $e_g$ electron at site $i$
occupies the orbital state
\begin{equation}
|\phi_i\rangle=\cos(\phi_i/2)d_{3z^2-r^2}
+\sin(\phi_i/2)d_{x^2-y^2}\ ,
\end{equation}
and it also represents
the direction of the lattice distortion; 
i.e, $\mbox{\boldmath{$\tau$}}_i
={\bf Q}_i/Q_i$. The effective hopping integral in Eq.\ (2)
can be written as  
\begin{equation}
\widetilde{t}_{ij}=t\sqrt{(1+{\bf s}_i\cdot{\bf s}_j)/2}
\sqrt{(1+\mbox{\boldmath{$\tau$}}_i\cdot{\bf n}_{ij}) 
(1+\mbox{\boldmath{$\tau$}}_j\cdot{\bf n}_{ij})}\ .
\end{equation}
The first square root together with $t$ is the well-known
hopping integral in the conventional DE model~\cite{b11},
and the last term describes the dependence on the orbital
alignments.

Equation (2) is a modified DE model. 
The main merit of this model
is that it contains only weak couplings between classical fields
${\bf s}_i$ and $\mbox{\boldmath{$\tau$}}_i$. 
To show its relevance to the Mn oxides, we shall
calculate the magnetic and orbital phase diagram.
Focusing on the spin and isospin dynamics, we replace 
the charge operators in Eq.\ (2) by their averages
$n_i\simeq n_e$ and $d_i^\dagger d_j\simeq 
\langle d_i^\dagger d_j\rangle$. We further approximate
$\langle d_i^\dagger d_j\rangle\simeq\langle d_i^\dagger d_j
\rangle_0$ with
$\langle d_i^\dagger d_j\rangle_0$ as the average 
of $d_i^\dagger d_j$ in 
a tight binding model.    
The resulting spin and isospin Hamiltonian is classic 
and can be studied by using Monte Carlo simulations together
with simulated
annealing technique~\cite{b18}. We start from a sufficiently high
temperature $k_BT=0.2 t$ and work on a $14\times 14
\times 14$ cubic lattice. The temperature is decreased 
by $8\%$ each step. For each step, about $2\times 10^{6}$
spin and isospin configurations are sampled, and thermal
average is calculated after initial sampling of 
$5\times 10^{5}$ configurations. 
 
We consider first the undoped case, where $n_e=1$
or $x=0$ and the kinetic energy term in 
Eq.\ (2) vanishes. The ground state depends 
on only two dimensionless
parameters $\lambda$ and $J'/J$ where $J=J_{\mbox{s}}
(1+\gamma)$. 
In the Monte Carlo simulation, as the temperature is
decreased to a sufficiently low value, e.g.,
$k_BT=10^{-2}J$, the spins and isospins
will eventually
develop their ground state configurations, 
which can be examined directly.
Numerical calculation indicates totally four possible 
ground states: G-AF, C-AF, A-AF and FM, where in the C-AF
state the local spins arrange ferromagnetically in parallel
chains and antiferromagnetically between neighboring chains. 
In all the four magnetic states,
the $e_g$ electrons occupy two different orbital states
on two sublattices, exhibiting alternating orbital
order. For the FM and G-AF states, the $e_g$
electrons occupy an arbitrary pair of orthogonal
orbital states in the two sublattices. For the
A-AF state, the occupied orbital states in the two
sublattices are $(d_{3z^2-r^2}+d_{x^2-y^2})/\sqrt{2}$
and $(d_{3z^2-r^2}-d_{x^2-y^2})/\sqrt{2}$. For the C-AF
state, the two occupied orbital states are
$d_{3x^2-r^2}$ and $d_{y^2-z^2}$, if the ferromagnetic
chains are assumed along the $y$ direction.
The obtained phase diagram in the $\lambda$ versus $J'/J$ plane
is given in Fig.\ 1. The phase transition between
different phases with changing the parameters 
is of first order.  

In order to ensure an A-AF ground state for undoped
$R$MnO$_3$,
the parameters have to be chosen in the shaded region
of Fig.\ 1. We will hereafter choose $J_{\mbox{s}}
=0.03t$, $U=42t$, $U'=7t$, $E_J=3.5t$ and $K=9t$
as reasonable values for the Mn oxides, which  
correspond to the point indicated  
by the triangle in Fig.\ 1 for the undoped case. 
The phase diagram in the normalized temperature 
$k_BT/t$ versus doping concentration $x$ plane is 
obtained numerically by calculating simultaneously 
various long-range order parameters such as the 
magnetization $M$ and  
short-range correlation functions of the localized
spins such as $\langle {\bf s}_i\cdot{\bf s}_j\rangle$
in three directions, as well as those of the isospins, as functions of 
$k_BT/t$ and $x$. The resulting phase diagram is shown in Fig.\ 2. 

In our phase diagram Fig.\ 2, there
are an A-AF, a G-AF and a FM 
phase at low temperatures
for $x\simeq 0$, $x\simeq 1.0$ and $
0.1\stackrel{<}{\sim}x\stackrel{<}{\sim}0.4$,
respectively, which are consistent with experiments
in most manganites~\cite{b1,b2,b3,b4,b5,bb5}. 
More importantly, we also observe an A-AF
state for $0.6\stackrel{<}{\sim}x\stackrel{<}{\sim}0.9$
and a first-order FM to A-AF phase transition 
upon cooling around $x\simeq 0.5$
[as will also be seen in Fig.\ (4)].
This result agrees well
with the experimental measurements in the manganites
with relatively wide conduction band, e.g., 
Pr$_{1-x}$Sr$_{x}$MnO$_3$~\cite{b5,b6,b7}. It is also 
possible, by slightly tunning the input parameters,
to make our phase diagram more comparable to
that of La$_{1-x}$Sr$_{x}$MnO$_{3}$, where a FM state
exists at $0.1\stackrel{<}{\sim}
x\stackrel{<}{\sim} 0.5$~\cite{bb5} and an A-AF state occurs for 
$x>0.52$~\cite{b10}. 

The magnetization $M$ normalized by the saturation
magnetization $M_{\mbox{s}}$ 
calculated using Monte Carlo technique  
is shown  in Fig.\ 3 as a function of the
normalized temperature 
for several values of doping concentration. 
For $x=0.1$, $0.2$ and $0.3$, the
magnetization approaches the saturation 
value $M_s$ at low temperatures, exhibiting 
strong ferromagnetism
in this doping range. The magnetization for $x=0.4$
drops sharply from a large value to zero at a low critical
temperature, corresponding
to a first-order FM to A-AF phase transition. 

Let us consider $x=0.5$. In Fig.\ 4, the magnetization
for different values of applied
 magnetic field is shown as a function of
the normalized temperature. At zero magnetic field
the magnetization first increases with decreasing temperature
and then drops abruptly, after which the system enters 
an A-AF state. This magnetization curve resembles
that measured in the Mn oxide Pr$_{0.5}$Sr$_{0.5}$MnO$_3$~\cite{b6}. 
With increasing the magnetic field, the magnetization increases
obviously. For $2\mu_BS^{*}H/t=0.024$, which corresponds to 
$H\simeq 20T$ if $t$ is taken to be $0.2$eV as a reasonable
value for the Mn oxides, the magnetization reaches about
$30\%$ even at very low temperatures. The sensitivity 
of the magnetization to the magnetic field 
may explain the sharp resistivity drop
induced by applied magnetic field observed in the low-temperature
A-AF phase of Pr$_{0.5}$Sr$_{0.5}$MnO$_3$~\cite{b6}.

Now we discuss the orbital structures in 
the phase diagram.  As seen from the inset
of Fig.\ 2, there are two different ground state orbital
structures. In the region denoted by ``antiparallel'',
the $e_g$ electrons occupy two different orbital states
$(d_{3z^2-r^2}+d_{x^2-y^2})/\sqrt{2}$ 
and $(d_{3z^2-r^2}-d_{x^2-y^2})/\sqrt{2}$
in two sublattices, forming alternating orbital order.
In the region denoted by ``parallel'', the $e_g$ electrons
occupy the same orbital $d_{x^2-y^2}$ on each site, 
showing uniform orbital order. The spatial distribution
of the charge density or the square of the orbital wave function
in the two types of orbital orders is illustrated in Fig.\ 5.
It is interesting to notice
that the antiparallel orbital order 
can be reguarded approximately
as between $d_{3y^2-r^2}$ 
and $d_{3x^2-r^2}$, because the overlaps between
these two sets of wave functions are 
larger than $97\%$. This result
is in good agreement with the orbital order
measured in LaMnO$_3$ using 
resonant X-ray scattering~\cite{b19}. 
By substituting the orbital wave functions into 
the Hamiltonian Eq.\ (2), 
it is straightforward to obtain an anisotropic
DE Hamiltonian for the charge carriers    
\begin{equation}
H_{eff}=-t\sum\limits_{ij}g_{ij}\sqrt{(1+{\bf s}_i\cdot
{\bf s}_j)/2}\;d_{i}^\dagger d_{j}\ ,
\end{equation}
where $g_{ij}$ assumes 
different values $g_\parallel$ and $g_\perp$
in the $ab$ plane and along the $c$
axis. For the antiparallel orbital order,
$g_\parallel=1/2$ and $g_{\perp}=1$, the motion
of the charge carriers is   
three-dimensional. For the parallel orbital order, 
$g_\parallel=3/2$ and $g_{\perp}=0$, the charge motion 
is of two-dimensional nature. According to Eq.\ (5),
the charge band width is $W=8t$ 
for the antiparallel orbital order, and $W=12t$ for
the parallel orbital order, 
indicating that the kinetic energy term 
favors parallel orbital order. The kinetic energy will compete
with the effective spin-orbital coupling which is
favorable for antiparallel orbital order.  
The spin-orbital coupling is proportional to $n_in_j
\sim (1-x)^2$
according to Eq.\ (2).
For small $x$, the spin-orbital coupling dominates
and leads to antiparallel orbital order, in which
the charges can move in three dimensions.
This explains the appearance of the FM state
for $0.1\stackrel{<}{\sim}x\stackrel{<}{\sim}0.4$
in the phase diagram Fig.\ 2.
For large $x$, the spin-orbital coupling reduces, 
the kinetic energy dominates and  
leads to parallel orbital order. In this case, 
the charge hopping is confined in  
two-dimensional planes, 
and so is the DE ferromagnetic coupling.
The net interplane spin interaction then
comes only from the antiferromagnetic superexchange coupling.
As a result the A-AF order occurs.  
 
In summary, we have established a modified DE model
to describe the spin and orbital properties in
the Mn oxides. The obtained phase diagram is 
consistent with experimental observations in 
the Mn oxides with relatively large conduction
band width, such as
Pr$_{1-x}$Sr$_{x}$MnO$_3$ and La$_{1-x}$Sr$_x$MnO$_3$,
where the effect of long-range Coulomb interactions may be 
unimportant and the 
charge ordering is absent or rather weak. 
We expect that the present
model can be further extended to describe  
the CE-AF state with charge ordering observed
in systems with relatively narrow conduction band
by including long-range Coulomb interactions.

This work was supported by the Texas Center for Superconductivity
at the University of Houston through a grant from the state
of Texas, by Texas ARP No: 3652707, and by the Robert A.
Welch foundation.

\begin{figure}
\caption{Phase diagram at $x=0$ in the $\lambda$
vs $J'/J$ plane where $J=J_{\mbox{s}}(1+\gamma)$.}
\end{figure}
\begin{figure}
\caption{Phase diagram in the $k_BT/t$ vs $x$ plane
with PM as the paramagnetic phase.
The dashed line indicates the boundaries of the 
orbital phases. Inset is the orbital phase diagram,
where ``antiparallel'' stands for alternating
orbital order between $(d_{3z^2-r^2}+d_{x^2-y^2})/\sqrt{2}$
and $(d_{3z^2-r^2}-d_{x^2-y^2})/\sqrt{2}$, and
``parallel'' indicates uniform $d_{x^2-y^2}$
orbital order.} 
\end{figure}
\begin{figure}
\caption{Calculated magnetization curves for several values
of doping concentration $x$.}
\end{figure}
\begin{figure}
\caption{Calculated magnetization curves for $x=0.5$ 
at several values of magnetic field.}
\end{figure}
\begin{figure}
\caption{Illustration of the spatial electron density distribution
in (a) alternating orbital ordering state between $(d_{3z^2-r^2}
+d_{x^2-y^2})/\sqrt{2}$ and $(d_{3z^2-r^2}-d_{x^2-y^2})/\sqrt{2}$,
and (b) uniform $d_{x^2-y^2}$ orbital ordering state.} 
\end{figure}
\end{document}